\documentclass[a4paper,11pt]{article}
\pdfoutput=1 

\usepackage{jinstpub} 
\usepackage{subfig,booktabs}

\title{\boldmath The Drift Chamber of the MEG II experiment}

\author[d,1]{ G.F.~Tassielli\note{Corresponding author.}}
\author[e]{, A. M.~Baldini}  
\author[a,b]{, G.~Cavoto}
\author[e,f]{, F.~Cei}
\author[e]{, M.~Chiappini}
\author[a,b,1]{, G.~Chiarello}
\author[d]{, A.~Corvaglia}
\author[e,f]{, M.~Francesconi}
\author[e]{, L.~Galli}
\author[d]{, F.~Grancagnolo}
\author[e]{, M.~Grassi}
\author[g]{, M.~Hildebrandt}
\author[a,b]{, M.~Meucci}
\author[d]{, A.~Miccoli}
\author[e,f]{, D.~Nicolò}
\author[c,d]{, M.~Panareo}
\author[e,f,g]{, A.~Papa}
\author[e]{, F.~Raffaelli}
\author[b]{, F.~Renga}
\author[g,h]{, P. Schwendimann}
\author[e]{, G.~Signorelli}
\author[b]{, C.~Voena}

\affiliation[a]{Dipartimento di Fisica dell'Università "La Sapienza",  Piazzale A. Moro, 00185 Roma, Italy}
\affiliation[b]{INFN Sezione di Roma, Piazzale A. Moro, 00185 Roma, Italy}
\affiliation[c]{Dipartimento di Matematica e Fisica dell’Università del Salento, Via per Arnesano, 73100 Lecce, Italy}
\affiliation[d]{INFN Sezione di Lecce, Via per Arnesano, 73100 Lecce, Italy}
\affiliation[e]{INFN Sezione di Pisa, Largo B. Pontecorvo 3, 56127, Pisa, Italy}
\affiliation[f]{Dipartimento di Fisica dell'Università di Pisa, Largo B. Pontecorvo 3, 56127 Pisa, Italy}
\affiliation[g]{Paul Scherrer Institut PSI, 5232 Villigen, Switzerland}
\affiliation[h]{ETH Z$\ddot u$rich, R$\ddot a$mistrasse 101, 8092, Z$\ddot u$rich, Switzerland}

\emailAdd{giovanni.tassielli@le.infn.it}

\abstract{The MEG experiment at the Paul Scherrer Institut searches for the charged-Lepton-Flavor-Violating $\mu^{+}\rightarrow e^{+}\gamma$ decay. MEG has already set the world best upper limit on the branching ratio: BR$<$4.2$\times10^{-13}$ @ 90\% C.l. An upgrade (MEG II) of the whole detector has been approved to obtain a substantial increase of sensitivity. Currently MEG II is completing the upgrade of the various detectors, an engineering run and a pre-commissioning run were carried out during 2018 and 2019.\\
The new positron tracker is a unique volume, ultra-light He based cylindrical drift chamber (CDCH), with high granularity: 9 layers of 192 square drift cells, $\sim$6-9 mm wide, consist of $\sim$12000 wires in a full stereo configuration. To ensure the electrostatic stability of the drift cells a new wiring strategy should be developed due to the high wire density (12 wires/cm$^{2}$), the stringent precision requirements on the wire position and uniformity of the wire mechanical tension (better than $0.5~g$). The basic idea is to create multiwire frames, by soldering a set of (16 or 32) wires on $40~\mu$m thick custom wire-PCBs. Multiwire frames and PEEK spacers are overlapped alternately along the radius, to set the proper cell width, in each of the twelve sectors defined by the spokes of the rudder wheel shaped end-plates. Despite to the conceptual simplicity of the assembling strategies, the building of the multiwire frames, with the set requirements, imposes a use of an automatic wiring system.\\
	The MEG II CDCH is the first cylindrical drift chamber ever designed and built in a modular way and it will allow to track positrons, with a momentum greater than 45 MeV/c, with high efficiency by using a very small amount of material, 1.5x10$^{-3}$ X$_{0}$. We describe the CDCH design and construction, the wiring phase at INFN-Lecce, the choice of the wires, their mechanical properties, the assembly and sealing at INFN-Pisa and the preliminary commissioning activities at PSI.}


\arxivnumber{2006.02378} 

\proceeding{International Conference on Instrumentation for Colliding Beam Physics\\
Budker Institute of Nuclear Physics, Novosibirsk, Russia\\
24 – 28 February 2020}

\begin{document}
\maketitle
\flushbottom

\section{Introduction}
\label{sec:intro}
\begin{figure}[ht]
\centering 
\begin{minipage}{0.4\textwidth}
\includegraphics[width=\textwidth]{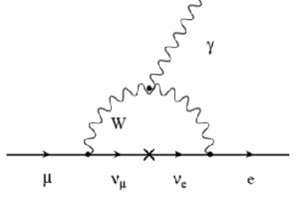}
\vspace{0.0cm}
\end{minipage}
\begin{minipage}{0.5\textwidth}
\centering
\includegraphics[width=0.8\textwidth]{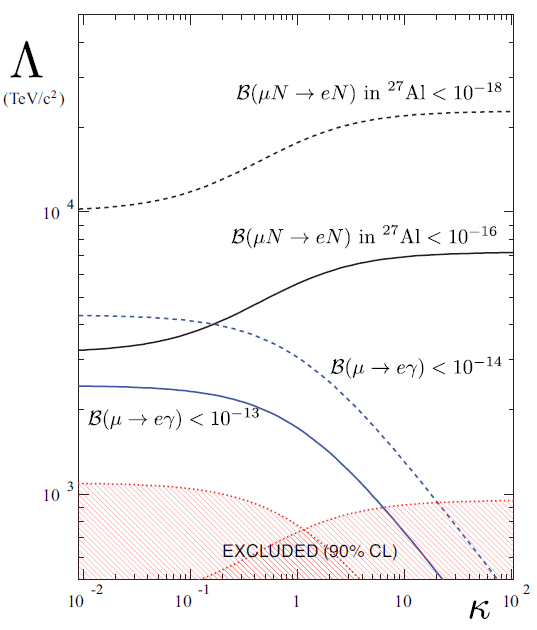}
\end{minipage}
\caption{\label{fig:clfv} Left: Diagram contributing to $\mu\rightarrow e\gamma$ in the SM with massive neutrinos. Right: Sensitivity of cLFV search with $\mu$ to the new physics mass scale $\Lambda$ as a function of $k$ \cite{Gou}, as defined in Eqn. \eqref{eq:lag}. The currently excluded regions of this parameter space from MEG and SINDRUM-II experiments are indicated. Comparisons of the sensitivity of the $\mu\rightarrow e\gamma$ searches with the $\mu\rightarrow e$ in $^{27}$Al searches.}
\end{figure}

Charged-Lepton-Flavour-Violation (cLFV) represents one of the best test beds for some of the open questions of the particle physics Standard Model (SM). After discovering of neutrino oscillations and setting the neutrino mass limits, the cLFV processes are allowed in the framework of the SM, see Fig. \ref{fig:clfv}-left, but at immeasurable levels, e.g. with a branching ratio BR($\mu\rightarrow e\gamma$) $\sim10^{-54}$ \cite{Mor}. However, many NP models predict cLFV processes at levels accessible by modern experiments. A detailed review is available in \cite{CSign}. Independently of the NP specific model, it is possible to assume a generic parameterization and an effective field theory (EFT) Lagrangian of the cLFV processes, roughly as the sum of a contribution related to the loop diagrams and another one related to the contact diagrams \cite{Gou}:
\begin{equation}
\label{eq:lag}
\mathcal{L}_{cLFV}=\frac{m_{\mu}}{(k+1)\Lambda^2}\overline{\mu}_{R}\sigma_{\mu\nu}e_{L}F^{\mu\nu} + 
\frac{k}{(k+1)\Lambda^2}\overline{\mu}_{L}\gamma_{\mu}e_{L}(\overline{f}\gamma^{\mu}f) + h.c.
\end{equation}
In Eq. \eqref{eq:lag} $\Lambda$ parameterizes the mass scale of the NP, $k$ is a dimensionless parameter that mediates between the two contributions, $m_{\mu}$ is the muon mass, $\mu/e/f$ are the Standard Model fermion fields and the subscripted L and R indicate their chirality, $F^{\mu\nu}$ is the photon field.
As a consequence, the relation between the BR and the sensitivity to a given mass scale range is different for every possible cLFV process, like radiative decays, non-radiative decays or direct muon to electron conversion in the Coulomb field of a nucleus. E.g. the $\mu\rightarrow e\gamma$ is mostly sensible to the loop interactions ($k\ll1$). It is worth noticing that the improvement of the BR upper limits on these processes imposes stringent constraints on the nature and on the scale $\Lambda$ of the possible NP, up to masses of the order of thousands of TeV, not accessible to direct searches. Fig. \ref{fig:clfv}-right shows the regions of $\Lambda-k$ parameter space excluded by measured and foreseen (during the next years) BRs for the different decays.
In this context the MEG experiment, at Paul Scherrer Institute (PSI), set the world best upper limit,  BR($\mu^{+}\rightarrow e^{+}\gamma$) $<4.2 \times 10^{-13}$ (90\% C.L.), exploiting the full statistics collected during the 2009-2013 data taking period \cite{MegI}. Moreover, in the next few years, the upgrade of the MEG experiment, MEG II, will investigate the $\mu^{+}\rightarrow e^{+}\gamma$ decay down to a BR of $6 \times 10^{-14}$.

\section{MEG II experiment}
\label{sec:megii}
\begin{figure}[t]
\centering 
\includegraphics[width=0.85\textwidth]{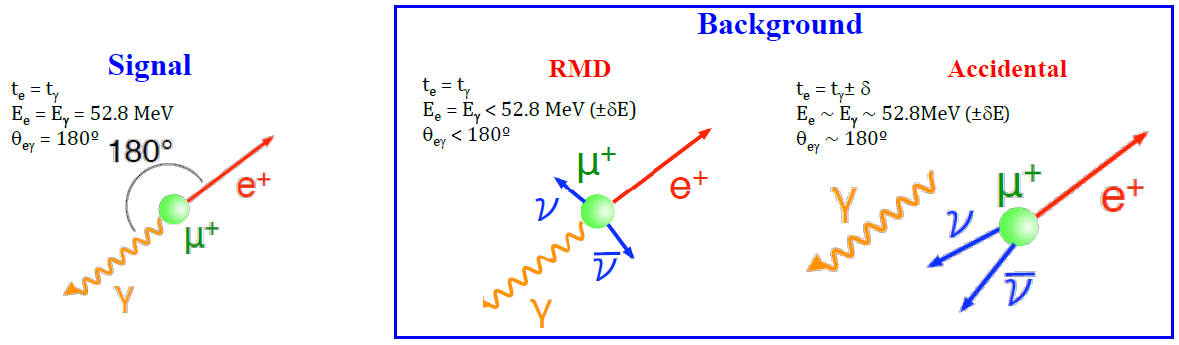}
\caption{\label{fig:megsgn} Left: $\mu\rightarrow e\gamma$ signature. Right: background sources: Radiative Muon Decay (RMD) and Accidental events overlap ($e$ and $\gamma$ from different sources, e.g. $e^{+}$ from Michel decay and $\gamma$ from RMD or $e^{+}e^{-}$ annihilation, with compatible kinematics to the $\mu\rightarrow e\gamma$).}
\end{figure}

\begin{figure}[t]
\centering
\includegraphics[width=1.0\textwidth]{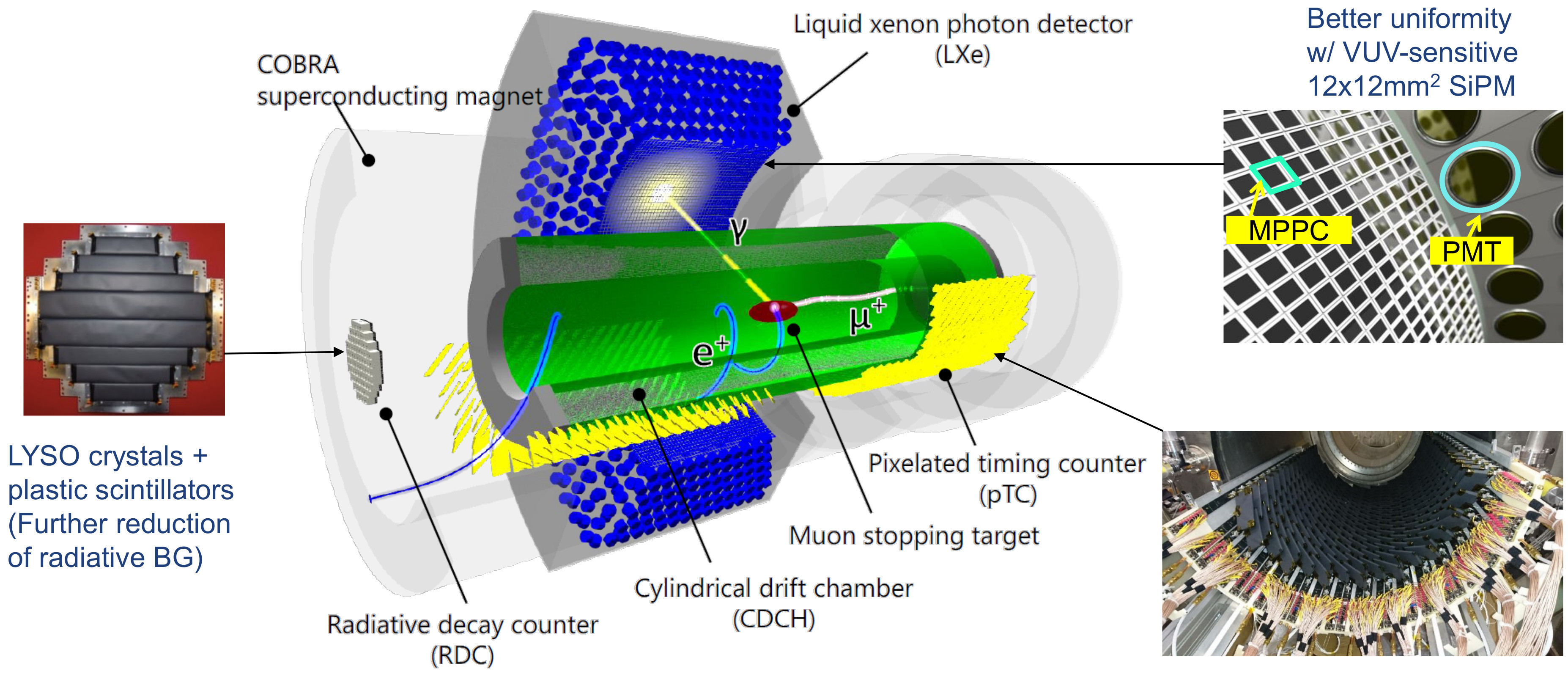}
\caption{\label{fig:megIIvw} Pictorial view of the MEG-II experiment with picture of some of the installed detectors: Radiative Decay Counter, Pixelated timing counter and the inner face of the Liquid Xe Calorimeter.}
\end{figure}

The MEG II experiment will search for the $\mu^{+}\rightarrow e^{+}\gamma$ decay by stopping a continuous surface $\mu^{+}$ beam, with a rate of $\Gamma_{\mu}=7 \times 10^{7} \mu$/s, on a 140 $\mu$m thin polyethylene target.
The search will be performed by looking for signal events made by an $e^{+}$ and a $\gamma$ in coincidence, collinear back-to-back and both with energies equal to one half of the muon mass (52.8 MeV). The background events are due to (see Fig.\ref{fig:megsgn}): a) the radiative muon decays (RMD, $\mu^{+}\rightarrow e^{+}\gamma\nu_{e}\overline{\nu}_{\mu}$) $N_{\rm RMD}$, where the neutrinos carry away little energy and the $e^{+}$ and the $\gamma$ are emitted back-to-back within the experimental accuracy; b) the accidental coincidence $N_{\rm acc}$ of an $e^{+}$ from a Michel decay ($\mu^{+}\rightarrow e^{+}\nu_{e}\overline{\nu}_{\mu}$) with a high energy $\gamma$ coming from either an uncorrelated RMD, or from an $e^{+}e^{-}$ annihilation-in-flight (AIF), or from the bremsstrahlung of an $e^{+}$ from a different Michel decay.
The total background is dominated by $N_{\rm acc}$, 
whereas $N_\text{RMD}$, proportional to $\Gamma_\mu$, is negligible with respect to it ($N_\text{RMD} \approx 0.1 N_\text{acc}$) \cite{MegI}. The number of signal events, $N_{\rm sig}$, and of the accidental background events, $N_{\rm acc}$, can be evaluated according to the following relations \cite{Kun}:
\begin{align}
N_{\rm sig}&=\Gamma_{\mu} \cdot T \cdot \Omega \cdot BR \cdot \varepsilon_{\gamma} \cdot \varepsilon_{e^{+}} \cdot \varepsilon_{s} \label{eq:nsig} \\
N_{\rm acc}&\propto\Gamma_{\mu}^{2} \cdot \Delta E_{\gamma}^{2} \cdot \Delta p_{e^{+}} \cdot \Delta \Theta_{e^{+}\gamma}^{2} \cdot \Delta t_{e^{+}\gamma} \cdot T \label{eq:nacc}
\end{align}
where the $\Delta E_{\gamma}$ is the photon energy resolution, $\Delta p_{e^{+}}$ the positron momentum resolution, $\Delta \Theta_{e^{+}\gamma}$ the resolution in the angle between the positron track and the extrapolated direction of the photon impact point to the stopping target, $\Delta t_{e^{+}\gamma}$ the relative time resolution between the photon and the positron, $T$ the total integrated acquisition time and $\Omega$ the geometrical acceptance. $\varepsilon_{\gamma}$, $\varepsilon_{e^{+}}$ and $\varepsilon_{s}$ are, respectively, the photon, the positron and the event selection efficiencies. 
The Eqs. \ref{eq:nsig}, \ref{eq:nacc} and a careful reassessment of the MEG apparatus with the modern detection techniques lead to the design of the upgrade of the experiment, MEG II, to reach an improvement of one order of magnitude with a reasonable and affordable costs and efforts. A detailed description of the MEG II experiment is reported in \cite{desMegII}, and the comparison between the MEG I and MEG II performances is reported in Tab. \ref{tab:megper}.
\begin{table}
\centering
\scalebox{0.8}{
\begin{tabular}{p{0.2\textwidth}p{0.16\textwidth}p{0.16\textwidth}p{0.16\textwidth}}
\toprule
Variable & Design MEG & Obtained MEG & Design MEG II \\
\midrule
$\Delta E_{e}$ (keV) & 200 & 380 & $\sim$100 \\
$\Delta \Theta_{e}, \Delta\phi_{e}$ (mrad) & 5, 5 & 9, 9 & 6, 5.5 \\
$Efficiency_{e}$ (\%) & 90 & 40 & >65 \\
$\Delta E_{\gamma}$ (MeV) & 1.2 & 1.7 & 1.0 \\
$\Delta Position_{\gamma}$ (mm) & 4 & 5 & <3 \\
$\Delta t_{e\gamma}$ (ps) & 65 & 120 & 85 \\
$Efficiency_{\gamma}$ (\%) & >40 & 60 & 70 \\
\bottomrule
\end{tabular}
}
\caption{\label{tab:megper} Comparison between MEG I (design and reached) performance and MEG II design ones.} 
\end{table}
\\In Fig. \ref{fig:megIIvw} a view of the MEG II experiment is reported with the highlights of some of the detectors upgrades, apart from the central Drift Chamber that is discussed in the next section.
The LXe is a homogeneous calorimeter able to fully contain the shower induced by a 52.83 MeV photon and to measure the energy, the interaction vertex and the interaction time with high efficiency. 
Its fiducial volume corresponds to $\sim$800 $l$ of LXe, covering 11\% of the solid angle viewed from the centre of the stopping target. In MEG the calorimeter was read out by 846 PMTs directly immersed in LXe. In MEG-II, in order to solve the issue of a non-uniform response due to the geometrical segmentation of the PMTs on the inner face, the 216 PMTs of the inner face have been replaced by 4092 vacuum UV (VUV) sensitive Multi-Pixel Photon Counters (MPPC) (6$\times$6 mm$^{2}$). The LXe calorimeter performance reached in MEG is: energy resolution of $\sim$1.7\% at 53 MeV, time resolution $\sim$64 ps and position resolution $\sim$5 mm. For MEG-II these resolutions have been improved and are expected to be $\sim$1.1\%, $\sim$50 ps and $\sim$3 mm, respectively.\\
The positron spectrometer consists of the COBRA (constant bending radius) magnet \cite{cobra} and a low-mass tracking system, composed of a drift chamber and a timing counter based on plastic scintillators. The COBRA magnet unchanged from MEG is a thin-walled, superconducting magnet with an axially graded magnetic field ranging from 1.27~T, at the centre of its axis, to 0.49~T, at both ends of the magnet cryostat. The graded field has the advantage, with respect to a uniform solenoid field, that particles produced with small longitudinal momentum have a much shorter transit time in the spectrometer, thus allowing for more stable operation in a high-rate environment. Moreover, the graded magnetic field is designed in such a way that positrons emitted from the target will follow a helical trajectory with an almost constant projected bending radius, only weakly dependent on the emission polar angle. The MEG II timing counter is made of 512 small scintillation tiles, it is named pixelated Timing Counter (pTC) and it replaces the MEG timing counter made by 30 scintillator (80 cm long) bars. The scintillation tiles (BC-422 scintillator with dimensions of 120$\times$(40 or 50)$\times$5 mm$^{3}$) are arranged in two semi-cylindrical super-modules placed symmetrically upstream and downstream in the spectrometer. Each tile is readout from two sides by 6 SiPMs connected in series for each side. The SiPMs are sensible to the near-ultraviolet (NUV), are 3$\times$3 mm$^{2}$ large with cells of 50$\times$50 $\mu$m$^{2}$. The whole timing counter has a total longitudinal and $\phi$ coverages of 23.0<$|z|$<116.7 cm and $-165.8^{\circ}$<$\phi$<$+5.2^{\circ}$ to fully cover the angular acceptance of the $e^{+}$ from $\mu^{+}\rightarrow e^{+}\gamma$ decays when the photon points to the LXe calorimeter. With this high granularity design the MEG II pTC has proven to reach the time resolutions of $\lesssim$100 ps per single counter and of $\lesssim$40 ps per positron track \cite{megIIptc}.\\
In MEG II an additional detector, the Radiative Decay Counter (RDC), is installed downstream to reduce the accidental background contribution and to increase the experiment's sensitivity by $\sim$16\%. It is used to identify a fraction of the low-energy $e^{+}$ from RMD decays having photon energies close to the kinematic limit, which are the main source of $\gamma$ for the accidental coincidence background. In the front part of the RDC there are 12 plastic scintillators, similar to the pTC ones, to measure the arrival times of $e^{+}$ with a resolution of $\sim$90 ps. Behind the plastic scintillators, there are 76 LYSO crystals with a size of 2$\times$2$\times$2 cm$^{3}$ each one readout by a SiPM (3$\times$3 mm$^{2}$). The RDC has proven to work in high rate environment ($\sim$MHz) and to detect the low energy $e^{+}$ with an adequate energy resolution ($\sim$6\%  at 1 MeV) to efficiently reject the accidental photons \cite{megIIrdc}.

\section{MEG II Cylindrical Drift Chamber}
\label{sec:cdch}
\begin{figure}
\begin{minipage}{0.62\textwidth}
\includegraphics[width=\textwidth]{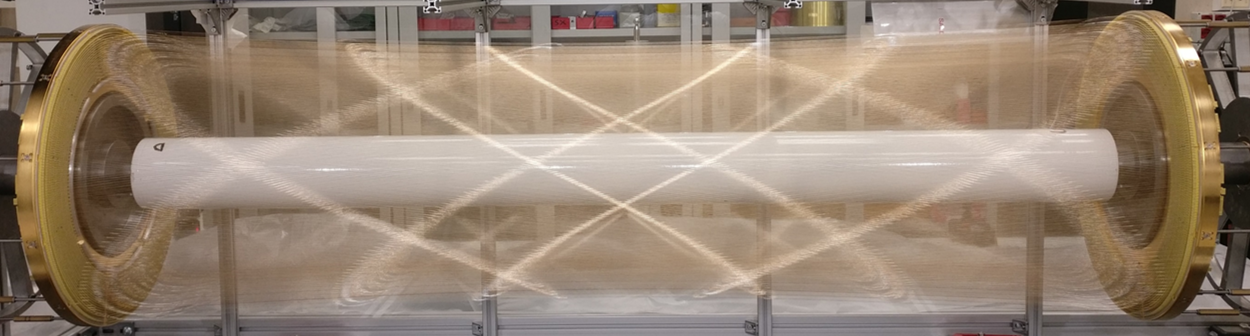}
\end{minipage}
\begin{minipage}{0.38\textwidth}
\includegraphics[width=\textwidth]{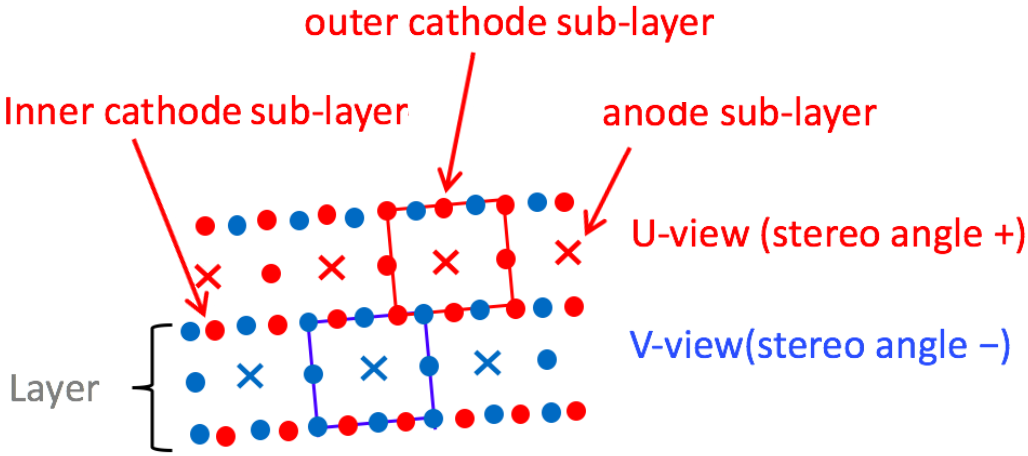}
\end{minipage}
\caption{\label{fig:cdch} Left: The MEG II CDCH fully wired on the assembly structure. Right: Its drift cells layout.}
\end{figure}

The MEG II positron tracker is a unique volume low-mass Cylindrical Drift CHamber (CDCH) with a high granularity and a full stereo wires configuration, as seen in Fig. \ref{fig:cdch}-left. The CDCH radially extends from R$_{\rm inner}$= 196 mm to R$_{\rm outer}$= 284 mm and the active area is $\sim$193 cm long. The new tracking system is designed to minimize the material budget crossed by positrons with momentum >45 MeV/c before reaching the pTC, $\sim$1.5$\times10^{-3} X_{0}$ per track turn. Moreover, it is designed to extend the tracker active area as close as possible to the pTC and to minimize the dead material between the two detectors. This feature allows to maximize the signal $e^{+}$ tracks reconstruction efficiency and the matching efficiency between CDCH and pTC tracks.\\
The CDCH is composed of 9 layers of drift cells at alternating stereo angles, divided in 12 identical 30$^{\circ}$ sectors per layer. Each layer has 192 square cells with a ratio of field to sense wires equal to 5:1 to ensure the proper electrostatic configuration, each layer is composed by one anode sublayer and two cathode sublayers, as sketched in Fig. \ref{fig:cdch}-right.
The cell size varies linearly between 6.7 mm, at the inner radius, and 8.7 mm at the outer radius. The alternating stereo angles vary from 5.8$^{\circ}$ to 8.5$^{\circ}$. The anodes are 20 $\mu$m diameter tungsten wires, while the cathode are 40, 50 $\mu$m aluminum wires. The 50 $\mu$m aluminum wires are placed on the sense wires layers, between two adjacent ones. The 40 $\mu$m aluminum wires are placed on the top and bottom sublayers and are placed with both the stereo view orientations per sub-layer. To equalize the gain of the innermost and outermost layers, two layers of guard wires, made with 50 $\mu$m aluminum wires, surround the active volume of the chamber. In total, the CDCH is made with 1728 sense wires, 9408 cathode wires and 768 guard wires. The CDCH volume is defined by a 2 mm thick carbon fiber shell at the outer radius and by a thin ( 20 $\mu$m) aluminized mylar foil at the inner radius. The external carbon fiber shell has the function of gas containment and of structural function, all the mechanical strength needed to tense the wires ($\sim$300 kg) is absorbed by it. The inner mylar foil serves to separate the drift gas from the pure He atmosphere that fills the inner volume where the stopping target is located. To avoid to damage the mylar foil, the gas inside the inner volume and the drift gas inside the CDCH volume are flowed maintaining constantly the same pressure between the two volumes by the same gas system, a detailed description is reported here \cite{megIIgas}. The CDCH drift gas is a 90:10 He:i-C$_{4}$H$_{10}$ mixture, chosen for its low radiation length, its fast enough drift velocity and the spatial resolution obtainable (respectively of $\sim$1400 m, $\sim$2 cm$/\mu$s and $\lesssim$120 $\mu$m, as measured with prototypes \cite{megIIres}). Due to the low ionization statistics of the He based gas mixture, the design gas gain (G) to efficiently operate the CDCH is set to G=5$\times$10$^{5}$.\\
The CDCH is designed to operate with a particle flux of $\sim$30 kHz/cm$^{2}$ at the inner layers and reach an overall, tracking and pTC matching, efficiency of $\gtrsim$65\%, a momentum resolution of $\sim$100 keV/c for 52.8 MeV positrons and angular vertexing resolution of $\sim$6 mrad. Detailed full simulation studies prove that the CDCH can meet the goals.\\
The CDCH is read out from both the ends of each sense wire, for a total of 3456 channels. The read-out is based on a low noise, low distortion and wide bandwidth differential pre-amplification stage, to work as front-end (FE), followed by the waveform digitizer, the WaveDREAM board \cite{megIIdaq} (with programmable sampling speed up to 2 GSPs and with an analogue bandwidth of 1 GHz). 432 FE boards are used to amplify and transmit, over a 5m long differential cable, the analogue signal coming from the chamber with overall gain $\sim$10 and bandwidth of $\gtrsim$700 MHz. Each FE board hosts 8 double stage pre-amplification channels in an area of$\sim$5.5$\times$8 cm$^{2}$, moreover they serve to delivery the HV to each drift cells. The current consumption for each channel is 60 mA at a voltage supply of $\pm$2.5 V, for a total power consumption of $\sim$300W per end-plate, therefore an appropriate cooling system based on both the recirculation of a coolant fluid and forced dry air cooling is used. More details about the FE electronics are in \cite{megIIfe1, megIIfe2}.\\
Even if the single cell spatial resolution is adequate to meet the MEG II requirements, the read-out allows the possibility to use the Cluster Counting technique \cite{impres, clco} to improve the spatial resolution and the CDCH performance, MEG II will explore this option.

\subsection{Construction technique of a high granularity and a high transparency drift chambers}
\begin{figure}
\includegraphics[width=\textwidth]{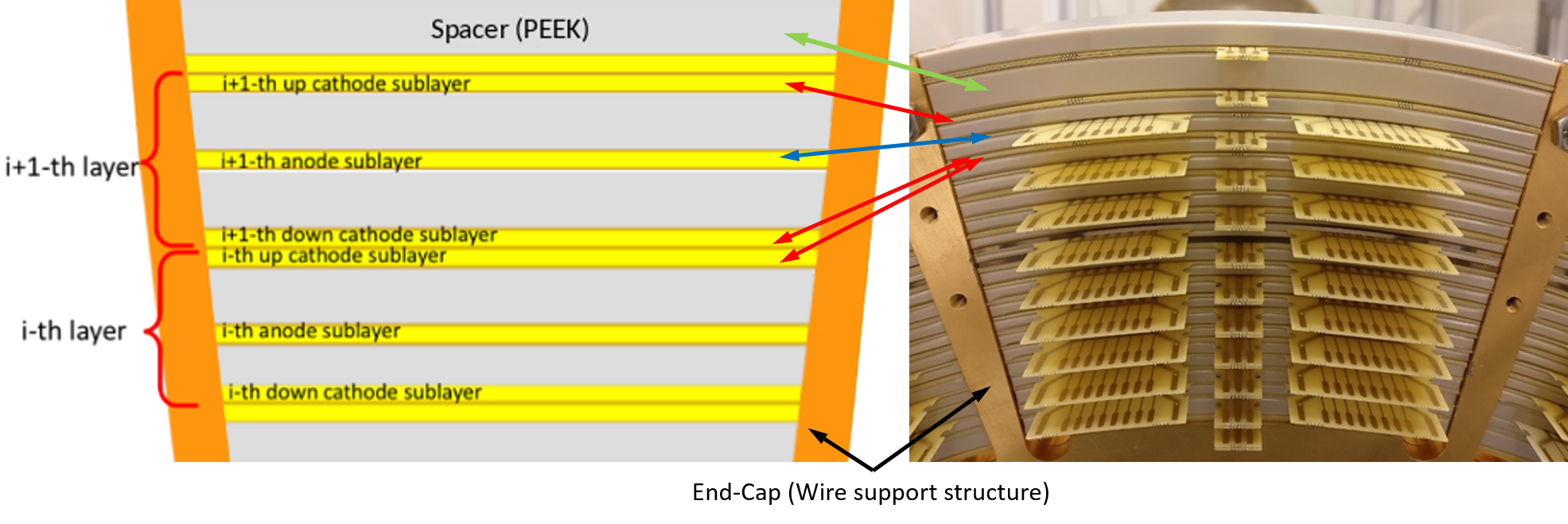}
\caption{\label{fig:cdchlay} Schematic view (left) and real picture (right) of the stack of wire PCBs and spacers on the end-cap
wiring support structure defining the drift chamber geometry.}
\end{figure}

Due to the high wire density, about 12~wires$/$cm$^{2}$, the use of the traditional feed-through technique as wire anchoring system cannot be implemented and therefore it was necessary to develop new wiring strategies. For the construction of the CDCH, MEG II followed the approach of separating the wire anchoring function by the mechanical and gas containing ones \cite{ulwmdch, nwctq}. The basic idea is to create a frame of multiwire (32 wires for the MEG II case), by soldering the wires between two 400 $\mu$m thick ad hoc designed Printed Circuit Boards (called wire PCBs). 
The chamber is assembled by stacking alternately a multi-wire frame and a spacer made of peek on the wire support end-plate, see Fig \ref{fig:cdchlay}.
Since the wires are just soldered on the PCBs, there is no limitation due to the physical dimensions of the feed-through, and the wires can be placed very close together. Despite of the conceptual simplicity of the building strategies, to ensure the electrostatic stability of the drift cells, great precision is needed in the following aspects: the end-plates and the spacers have to be numerically machined to guarantee the mechanical precision in the wire PCBs positioning; the wires must be placed with an accuracy of 20 $\mu$m and stretched with an accuracy of $\lesssim$0.5 g. This means that a wiring robot must be used to produce all the 396 (plus spares) multi-wire frames with the necessary accuracy and homogeneity, as described below.

\subsection{The wiring robot and the stringing procedures}
\begin{figure}[t]
\begin{minipage}{0.5\textwidth}
\includegraphics[width=\textwidth]{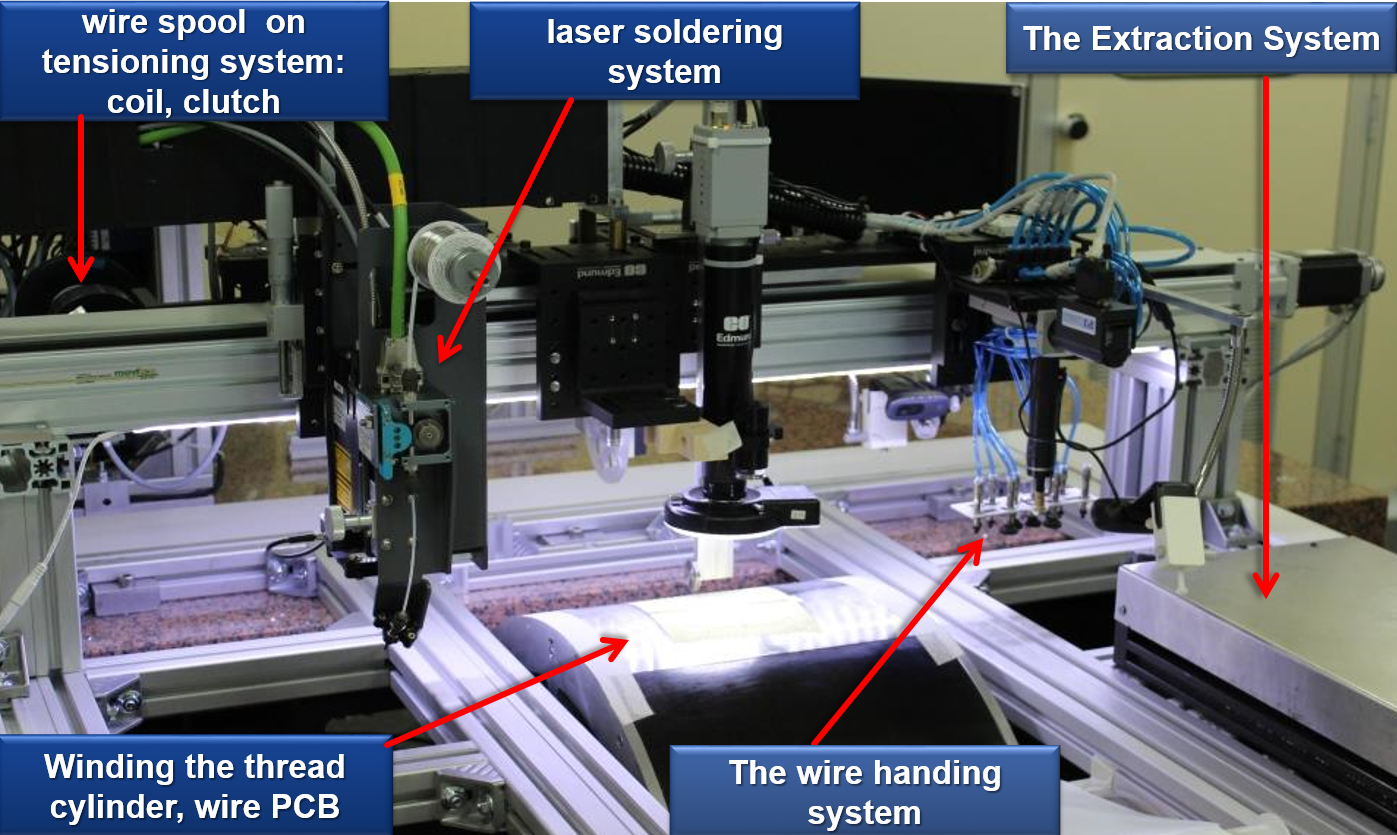}
\end{minipage}
\begin{minipage}{0.5\textwidth}
\centering
\includegraphics[width=\textwidth]{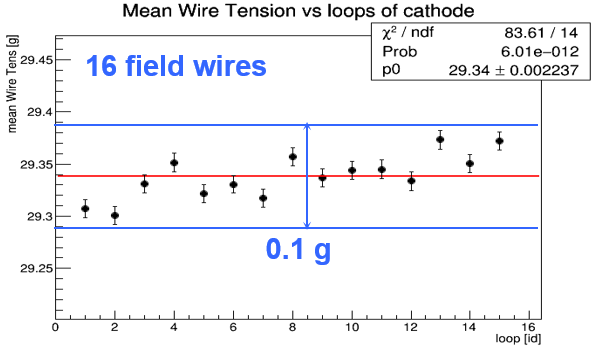}
\end{minipage}
\caption{\label{fig:wrrob} Left: MEG II CDCH wiring robot. Right: Average wire tension applied for each loop.}
\end{figure}

The wiring robot, shown in Fig. \ref{fig:wrrob}-left, has been designed for: monitoring the wire positions and their alignments within a few tens of $\mu$m; applying the wires mechanical tension and maintaining it constant and uniform through the whole chamber; soldering the wires on the wiring PCBs, by using a contactless procedure, and monitoring the solder quality. In order to fulfill all these requirements, the wiring robot is made up of three systems: the {\bf WIRING}, the {\bf SOLDERING} and the {\bf AUTOMATIC HANDLING}. A real-time system, based on a National Instrument Compact RIO platform\textregistered, controls the three systems simultaneously, sequencing and synchronizing all the different operations.
\\The purpose of the {\bf WIRING} system is the winding of a multi-wire layer. Two PCBs, aligned and oriented at the proper stereo angle, are placed back-to-back on the winding cylinder. The multi-wire layer is obtained in a single operation by winding along a helical path the same wire multiple times (32 for cathodic type layers and 2 times 16 for anodic type layers) around the cylinder with a pitch corresponding to the wire PCBs spacing. The mechanical tension to the wire is delivered by an electromagnetic brake, acting on the wire spool, and it is on-line monitored by a high precision dedicated system. The measured wire tension value is used in a real-time feedback system to correct the tension applied by the electromagnetic brake. Thanks to the feedback system the wire tension is delivered with an instantaneous variation of 0.15~g and a variations of less than 0.05~g per single wire during a wiring sequence, see Fig. \ref{fig:wrrob}-right.\\
The {\bf SOLDERING} system is based on an IR laser soldering station. To avoid damages to the frail Al field wires, the soldering process is done at a temperature of 180$^{\circ}$C by using a low melting point (160$^{\circ}$C) tin alloy. The laser system is controlled and synchronized with the positioning system that localizes automatically the soldering pad by using a pattern matching software developed with LabVIEW\textregistered.\\
The {\bf AUTOMATIC HANDLING} system unrolls and remove tension from the wound layer of soldered wires around the cylinder to storage them on a transport frame, this is accomplished with an automatic device. The first wire PCB is lifted off from the cylinder surface with a linear actuator connected to a set of vacuum-operated suction cups and then it is placed on the storage and transport frame. Unrolling is accomplished by synchronizing the cylinder rotation with the linear displacement of the frame. Once the layer of soldered wires is completely unrolled, the second wire PCB is lifted off from the cylinder, as the first one, and placed on the transport frame. More details about the wiring robot are reported in \cite{desMegII, autrob}.

\subsection{The assembly procedures}
39 built multi-wire layers were shipped, every three weeks, from INFN-Lecce to INFN-Pisa where the CDCH was assembled. The wire PCBs of the multi-wire planes were kept parallel in order to avoid stresses to the wires at the soldering points. For this reason the assembly procedure was performed with a DEA Ghibli coordinate measuring machine. The machine allows a position accuracy of about 20 $\mu$m in the horizontal plane and 40 $\mu$m on the vertical axis of the fiducial markers on the wire PCBs. After passing a mechanical stress test (a repeated cycle of elongation up to 25\% over the nominal tension), and a check of the wire tension by measuring their resonant frequency \cite{fullcdch}, the multi-wire frame is mounted on the chamber wire support end-plates. The wire support end-plates, Fig. \ref{fig:cdchlay}-right, have a helm shape with 12 spokes at 30$^{\circ}$, one per sector and are made of gold-plated aluminum with a thickness of 30 mm. They are loaded with a total force of 3000N due to the tension on the wires ($\sim$24.5 g for sense wires, $\sim$19.2 g and $\sim$29.6 g, respectively, for 40 $\mu$m and 50 $\mu$m field wires) and are designed to have a maximum deflection at the extremity of the helm spokes of about 200 $\mu$m, tolerable, given the wire elongation. During the assembly phase, these end-plates are moved to a shorter distance than the nominal one to avoid stressing the wires in the procedure. The mounting procedure was performed by using an adjustable arm to release a multi-wire frame from the transport support and, by placing it next to the end-plates for the engagement procedure, to transfer the multi-wire frame on the end plates between two spokes. The final positioning was driven by hand through dedicated nippers. The wire PCBs were glued on the peek spacer with a double-sided tape previously applied on the inner layer. This procedure was repeated for each of the 12 sectors and for the 9 layers. After mounting the outermost layer, the end-plates were moved to the nominal distance and the CDCH was closed with the outer structural carbon fiber cylinder. Then the end-plates were sealed to prevent gas leakage by using the ThreeBond 1530 glue and the Stycast 2850 resin, after the sealing the CDCH was shipped to PSI. Before the insertion of the CDCH into the experiment it was equipped with the FE electronics, their supports and cooling pipes.\\
During the assembly and commissioning phases a few ($\sim$65) Aluminum wires broke and they were safely extracted from the chamber. Absence of a few field wires will hardly affect the electrostatic configuration of the chamber, as confirmed by simulations. The broken wires were deeply investigated by performing optical inspections with microscopes, chromatography, practical tests and SEM/EDS analyses. Chemical and mechanical analyses showed that the origin of the breaking phenomenon is a chemical corrosion of the wire core in presence of water condensation on wires. It was discovered that the clean room atmosphere was not dry enough for the safety of the used Aluminum wires. Two hypotheses are identified as possible sources of the corrosion, an intermetallic galvanic process between the Al wire core and the Ag used for the wire coating or a corrosion by Cl contaminations (brought by water vapors) that can reach the Al wire core through imperfections of the wire coating. The tests performed were not able to distinguish between the two possible sources. The corrosion is stopped after the chamber volume was in an absolutely dry atmosphere with a continuous flow of inert gas (Nitrogen or Helium).

\subsection{Commissioning status}
\begin{figure}[t]
\begin{minipage}{0.478\textwidth}
\includegraphics[width=\textwidth]{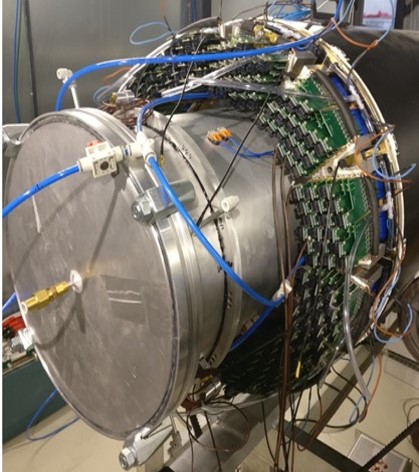}
\end{minipage}
\begin{minipage}{0.522\textwidth}
\centering
\includegraphics[width=\textwidth]{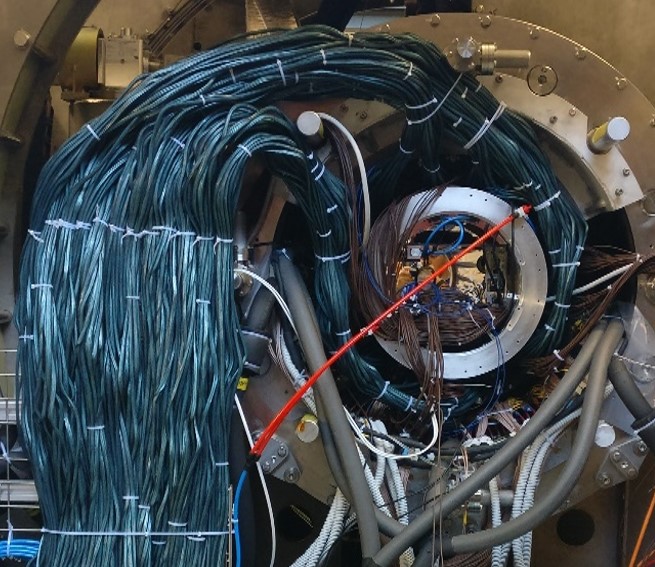}
\end{minipage}
\caption{\label{fig:CDCHin} Left: MEG II CDCH during the FE boards installation. Right: CDCH fully integrated into the MEG II apparatus, the signal cables together with the gas/cooling system pipes are visible.}
\end{figure}

\begin{figure}
\begin{minipage}{0.51\textwidth}
\includegraphics[width=\textwidth]{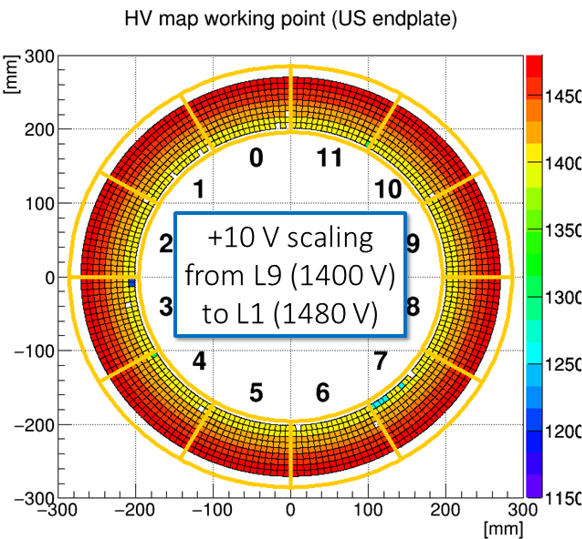}
\end{minipage}
\begin{minipage}{0.49\textwidth}
\centering
\includegraphics[width=\textwidth]{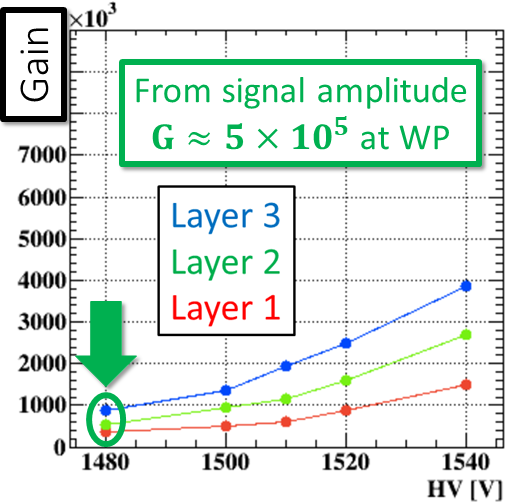}
\end{minipage}
\caption{\label{fig:CDCHwrk} Left: MEG II CDCH final HV map at the working point. The color scale ranges from 1150 V to 1480V. Right: Experimental gain curves as extracted from cosmic rays data.}
\end{figure}

The CDCH assembly was completed in 2018. Thereafter it was installed two times in the experiment and tested under MEG II beam conditions during the engineering run (winter 2018), and the first commissioning run (winter 2019), see Fig \ref{fig:CDCHin}.
 After the 2018 run, maintenance operations were performed, e.g. the broken wires removal, to fix the working condition of the chamber. Based on the average atmospheric pressure $\sim$970 mbar at PSI, the gas mixture and the cell dimension, the CDCH HV working point (WP) to reach the requested gas gain is at 1400 V for the innermost layer and at 1480 V for the outermost. It increases by 10 V for each radial layer. In Fig. \ref{fig:CDCHwrk}-left the HV map configuration reached by each single CDCH cell is reported. The map shows that a few of them ($\sim$1.3\% of all cells) don't reach the WP. However, the impact of this cells inefficiency on the reconstruction performance was proven to be negligible by the full simulation.
The electrostatic stability of the drift cells was tested up to a safety margin of additional 100 V above the WP. This is needed to have some margin to compensate the gain loss due to operation condition variation and to the expected chamber aging \cite{desMegII}.\\
During the runs only 192 DAQ channels were available, corresponding to six layers in one sector (16 drift cells per layer) with the double-side read out that, due to the full stereo configuration, did not allow to test the particle tracking. Nevertheless, we studied the noise level in the experimental environment and performed several HV scans around the WP with Cosmic Rays (CR) and with the $\mu^{+}$ beam. With CR data the first experimental gas gain studies was performed. In Fig. \ref{fig:CDCHwrk}-right the preliminary measurements of the gas gain are shown that are in agreement with the expectations. The muon beam was used to test the chamber response in a high rate environment and its synchronization with the other sub-detectors. Moreover, the cell occupancy and rate capability was studied as a function of the beam rate by running at three beam intensities 1, 3 and 7 $\times$10$^{7}$ $\mu^{+}$/s, showing good agreement with the expected chamber rate capability.\\
More details about the CDCH commissioning are reported in \cite{megIIcomm}.
 
\section{Summary}
\label{sec:summ}
The positron tracker of the MEG II experiment is a full stereo Cylindrical Drift Chamber, with high granularity and high transparency ($\sim$1.5$\times10^{-3} X_{0}$), built with an innovative construction technique. It is a crucial detector to improve the angular and momentum resolutions of the $e^{+}$ kinematic variables, in addition to a contribution to the minimization of the background sources. The CDCH mechanics proved to be stable and adequate to sustain a MEG II run, even if there were some ($\sim$60 over $\sim$12000) broken wires, all of them due to a corrosion issue. No additional wire breaking was found after the closing of the chamber in the right atmosphere.\\
During the commissioning activities at PSI the working configuration was reached and CDCH was fully integrated into the MEG II experimental apparatus for the data taking. Two first commissioning runs were performed during 2018 and 2019 and some tests and measurements on the gas gain and rate capability were done. We expect to reinstall the chamber in 2020 to finalize the commissioning.

%
%

\end{document}